\documentclass[fleqn,usenatbib]{mnras}

\usepackage{newtxtext,newtxmath}

\usepackage[T1]{fontenc}

\DeclareRobustCommand{\VAN}[3]{#2}
\let\VANthebibliography\thebibliography
\def\thebibliography{\DeclareRobustCommand{\VAN}[3]{##3}\VANthebibliography}

\usepackage{xcolor}

\usepackage{booktabs}
\usepackage{caption}
\usepackage{subcaption}

\usepackage{booktabs}

\usepackage{multirow}
\usepackage{booktabs}

\usepackage{graphicx}	
\usepackage{amsmath}	
\usepackage{url}

\newcommand{\hGpc}{\,h^{-1}~{\rm Gpc}}

\newcommand{\hMpc}{{\ifmmode{\,h^{-1}{\rm Mpc}}\else{$h^{-1}$Mpc}\fi}}
\newcommand{\hkpc}{{\ifmmode{\,h^{-1}{\rm kpc}}\else{$h^{-1}$kpc}\fi}}
\newcommand{\hMsun}{{\ifmmode{\,h^{-1}{\rm {M_{\odot}}}}\else{$h^{-1}{\rm{M_{\odot}}}$}\fi}}

\newcommand{\Mstar}{{\ifmmode{\,M_{*}}\else{$M_{*}$}\fi}}
\newcommand{\Mhalo}{{\ifmmode{\,M_{\rm halo}}\else{$M_{\rm halo}$}\fi}}
\newcommand{\ltsima}{$\; \buildrel < \over \sim \;$}
\newcommand{\gtsima}{$\; \buildrel > \over \sim \;$}
\newcommand{\lsim}{\lower.5ex\hbox{\ltsima}}
\newcommand{\gsim}{\lower.5ex\hbox{\gtsima}}

\newcommand{\gadgetx}{\textsc{Gadget-X}}

\newcommand{\gizmo}{\textsc{GIZMO-SIMBA}}

\newcommand{\thethreehundred}{{\sc The Three Hundred}}

\title[ML cluster mass radial profiles from SZ maps]{The Three Hundred project: A Machine Learning method to infer clusters of galaxies mass radial profiles from mock Sunyaev-Zel'dovich maps}

\author[A. Ferragamo et al.]{
A. Ferragamo,$^{1}$\thanks{E-mail: antonio.ferragamo@uniroma1.it}
D. de Andres,$^{2,3}$\thanks{E-mail: daniel.deandres@uam.es}
A. Sbriglio$^{1}$,
W. Cui$^{2,3,4}$,
M. De Petris$^{1}$,
G. Yepes$^{2,3}$,
R. Dupuis$^{5}$,
\newauthor
M. Jarraya$^{5}$,
I. Lahouli$^{5}$,
F. De Luca$^{6}$,
G. Gianfagna$^{1,7}$,
E. Rasia$^{8,9}$
\\
$^{1}$Dipartimento di Fisica, Sapienza Universit\`a di Roma,Piazzale Aldo Moro 5, I-00185 Roma, Italy\\
$^{2}$Departamento de F\'isica T\'eorica, M\'odulo 8, Facultad de Ciencias, Universidad Aut\'onoma de Madrid, E-28049 Madrid, Spain\\
$^{3}$Centro de Investigaci\'on Avanzado en F\'isica Fundamental (CIAFF), Facultad de Ciencias, Universidad Aut\'onoma de Madrid, E-28049 Madrid, Spain\\
$^{4}$Institute for Astronomy, University of Edinburgh, Edinburgh EH9 3HJ, United Kingdom \\
$^{5}$EURANOVA, Mont-Saint-Guibert, Belgium\\
$^{6}$Dipartimento di Fisica, Universit\`a di Roma Tor Vergata, Via della Ricerca Scientifica 1, I-00133 Roma, Italy \\
$^{7}$INAF - Istituto di Astrofisica e Planetologia Spaziali, via Fosso del Cavaliere 100, I-00133 Roma, Italy \\
$^{8}$IFPU - Institute for Fundamental Physics of the Universe, Via Beirut 2, I-34014 Trieste, Italy \\
$^{9}$INAF Osservatorio Astronomico di Trieste, via Tiepolo 11, I-34131, Trieste, Italy
}

\date{Accepted XXX. Received YYY; in original form ZZZ}

\pubyear{2022}

\begin{document}
\label{firstpage}
\pagerange{\pageref{firstpage}--\pageref{lastpage}}
\maketitle

\begin{abstract}

We develop a machine learning algorithm to infer the 3D cumulative radial profiles of total and gas mass in galaxy clusters from thermal Sunyaev-Zel'dovich effect maps. We generate around 73,000 mock images along various lines of sight using 2,522 simulated clusters from the \thethreehundred{} project at redshift $z< 0.12$ and train a model that combines an autoencoder and a random forest.
Without making any prior assumptions about the hydrostatic equilibrium of the clusters, the model is capable of reconstructing the total mass profile as well as the gas mass profile, which is responsible for the SZ effect. We show that the recovered profiles are unbiased with a scatter of about $10\%$, slightly increasing towards the core and the outskirts of the cluster. 
We selected clusters in the mass range of $10^{13.5} \leq M_{200} /(\hMsun) \leq 10^{15.5}$, spanning different dynamical states, from relaxed to disturbed halos. We verify that both the accuracy and precision of this method show  a slight dependence on the dynamical state, but not on the cluster mass.
To further verify the consistency of our model, we fit the inferred total mass profiles with an NFW model and contrast the concentration values with those of the true profiles.
We note that the inferred profiles are unbiased for higher concentration values, reproducing a trustworthy mass-concentration relation. 
The comparison with a widely used mass estimation technique, such as hydrostatic equilibrium, demonstrates that our method recovers the total mass that is not biased by non-thermal motions of the gas.

\end{abstract}




\section{Introduction}

Small density fluctuations in the early Universe were the seeds for structure formation. The latest stage of structure evolution is characterised by the formation of clusters of galaxies. Galaxy clusters are the largest gravitationally bound structures in the Universe, reaching a mass of a few $10^{15}\hMsun$. The majority of this mass, about 80\%, corresponds to dark matter (DM), 12\% is diffused hot gas, i.e. the Intra Cluster Medium (ICM) and the $8\%$ are galaxies within the ICM \citep[see][for a review]{Kravtsov2012}. The abundance of galaxy clusters as a function of the mass and redshift, i.e. the halo mass function, is crucial for constraining cosmological parameters \citep[e.g.][]{Allen2011, Pratt2019}.

However, the DM component cannot be directly observed. On the contrary, the baryons could be revealed in the optical band via star/galaxies emission, at infrared wavelengths with the dust emission, in the X-ray band via bremsstrahlung emission, or at sub/millimetre wavelengths via the Sunyaev-Zeldovich \citep[SZ, ][]{SZ1970} effect.
Through the information that the baryon component gives us, it is therefore possible to have some hints on the cluster's total mass.
Common approaches exploit: (1) X-ray and SZ observations to recover cluster potential well from the ICM distribution under the assumption of Hydrostatic Equilibrium (HE), (2) mapping background lensed objects revealing cluster mass distorting power, and (3) galaxy members kinematics indicating potential well, see \citet{Pratt2019} for a review.
Due to the basic assumptions and measurement difficulties of each of the methods, the inferred mass could be affected by biases. These biases are commonly quantified in hydrodynamical simulations comparing the true and method-derived masses.
In a recent review by \citet{Gianfagna2021}, the mass, estimated with the HE model in synthetic clusters, is from a minimum of $ 10 \% $ to a maximum of $ 20 \% $ lower than the real value. Consistent results are also derived in  clusters from \thethreehundred{} simulations \citep{Gianfagna2022}.
The origin of this bias is still not totally  constrained as well as its dependence on  the cluster properties, mainly the dynamical state, and  the redshift. Non-thermal pressure support  due to different gas motion components \citep{Lau2009} could have an impact on the cluster mass budget. Moreover, the bias seems to be affected by the  cluster relaxation state \citep{Ansarifard2020, Gianfagna2022}. More intriguing is the bias dependence on  the redshift. While simulations agree on a negligible dependence \citep{Henson2017, Gianfagna2022}, observational data support that the estimated  masses are more biased at higher redshift \citep{Sereno2017, Wicker2022} but probably this can be due to observational mass selection effect.

Therefore, even if the cluster total mass is a powerful  tool to constrain the  evolution of the Universe, inaccurate estimates of this quantity make a large  impact in the inference of the cosmological parameters \citep{Pratt2019, Salvati2020}.

Recently, Machine Learning (ML) models have started to be applied  for estimating  the  cluster mass from mock multi-band images. \citet{Ntampaka2015, ntampaka2016, ntampaka2017}. More recently \citet{Ho2019} suggested that  ML algorithms can also be used to assess the effect of interlopers in the dynamical cluster mass estimates.
Exploiting the power of Convolutional Neural Networks (CNN), this  methodology is  playing an important role in 
the analysis of synthetic  observations of  X-ray \citep{Ntampaka2019} and SZ \citep{Gupta2020, Gupta2021, Yan2020}. Moreover, it has been recently  applied for the first time, on real cluster observations from Planck Compton-$y$ parameter maps \citep{deandres2022Planck}. 
By considering no physical  assumptions about the gas in  clusters, this  technique can ideally infer unbiased mass  values for real clusters.

In this work, we present a  combination of deep learning architecture, followed by an ML regression method that has been developed in order to infer  the total mass radial profile of clusters extracted from \thethreehundred{} simulations. We take as inputs a large sample of  mock SZ maps quantified in terms of the Compton $y$-parameter. Moreover, in order to obtain an independent estimate of the gas fraction, we designed our machinery to infer simultaneously the cluster gas mass radial profile.

This paper is organised as follows: in Sec.~\ref{sec:dataset} we introduce the simulated dataset, based on a cluster sample extracted from \thethreehundred{} project, and the mock SZ maps. 

In Sec.~\ref{sec:method} we briefly describe the deep learning architecture, based on an autoencoder that is used to extract features from SZ maps and the random forest regression algorithm that performs the mass profiles inference.  The results of our study are reported in Sec.~\ref{sec:results} where we also analyse the performance of our method. Finally, in Section~\ref{sec:discussion} we compare our results with more classical approaches based on the HE approximation and we summarise our main conclusions in Section~\ref{sec:conlusions}. 

\section{Data set}
\label{sec:dataset}

\subsection{Simulation}
\label{subsec:sim}

This study is based on synthetic clusters generated in \thethreehundred{} project, firstly introduced in \citet{Cui2018}. This consists in a set of zoomed hydrodynamic simulations of 324 spherical regions of $15\,\hMpc$ radius centred on the most massive clusters ($M_{\rm vir}>8\times 10^{14}\,\hMsun$, at $z = 0$) identified  within the dark-matter-only MultiDark Planck 2 simulation \citep[MDPL2, ][]{Klypin2016} by the Rockstar  halo finder \citep{Behroozi2013}. The MDPL2 simulation is a comoving volume of $(1\hGpc)^3$ containing $3840^3$ DM particles of mass $1.5 \times 10^9\,\hMsun$ and implements the Planck cosmology \citep[$\Omega_m = 0.307$, $\Omega_b = 0.048$, $\Omega_\Lambda= 0.693$, $h = 0.678$ $\sigma_8 = 0.823$, $n_s = 0.96$, ][]{Planck_cosmopar2015}. To resimulate each of \thethreehundred{} region with the full baryonic physics, the particles within the sphere of radius $15\,\hMpc$ where mapped back to the initial conditions and were splitted into dark matter ($M_\text{DM} = 1.27 \times 10^9\,\hMsun$) and gas ($M_\text{gas} = 2.36 \times 10^8\,\hMsun$) particles according to the universal baryon fraction of the Universe as estimated by {\textit{Planck}}, preserving the original mass resolution. The remaining particles outside the zoomed regions were resampled as  low-resolution particles in order to take into account the large-scale gravitational tidal field and reduce the computational cost. The hydrodynamical resimulations were performed by using the TreePM+SPH \textsc{GADGET-X} code, a modified version of \textsc{GADGET3} code that includes an improved Smoothed-Particle Hydrodynamics (SPH) scheme to account for  the gas dynamics of the baryonic component in  the simulations \citep{Springel2005,Beck2016}.  The code also includes  metal-dependent cooling  as described in \citet{tornatore07}. Star formation and Supernovae heating are modelled following the scheme of \citet{SH03}. Moreover, the effects of AGN feedback via gas accretion onto supermassive black holes are also taken into account as described in  \citet{Steinborn15}.

Each of the 324 resimulated regions was then analysed by using the Amiga's Halo Finder \citep[AHF, ][]{AHF2009}. It detects all halos by identifying the local peaks in the total density field interpolated from particles onto  a hierarchical mesh structure. It then estimates $R_{200}$ of each halo, as the radius at which the density of the object  reaches $200$ times the critical density of the Universe $\rho_c$, and $M_{200}$, as the mass of all the particles dynamically bounded to the cluster that lie within this radius. Hereafter, we refer to  these quantities as $R_{200}^\text{true}$ and $M_{200}^\text{true}$.

\begin{figure}
	\includegraphics[width=\columnwidth]{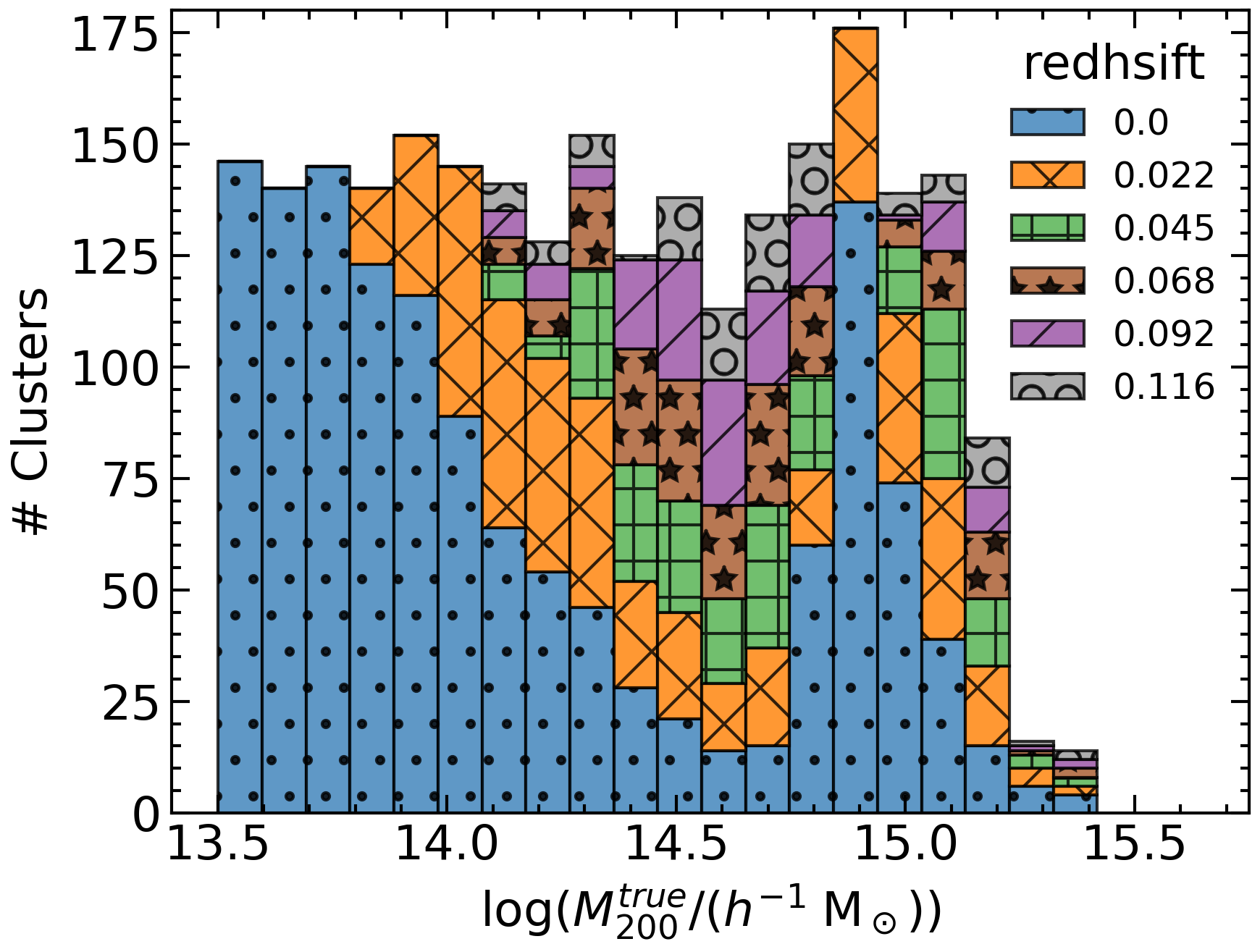}
    \caption{Mass distribution of clusters at the selected redshifts. the sample is first selected at redshift z=0 (blue bars), then it is complemented with clusters at the other redshifts (represented by the colours defined in the legend) to make it homogeneous.}
    \label{fig:sample_distribution}
\end{figure}

The sample used in this work is made of 2522 clusters uniformly distributed within the  mass interval $10^{13.5} \leq M_{200}/(\hMsun) \leq 10^{15.5}$ at six nearby redshifts, from $z=0$ to $z=0.116$, in order to have an almost homogeneously mass populated sample. Note however that there must be  always  fewer  halos in the larger mass range, see Fig.~\ref{fig:sample_distribution}.

\subsection{Cluster mass radial profiles}
\label{sec:data}

For each cluster, we extract the cumulative 3D radial profiles of the total mass and the gas mass. These profiles are obtained by summing up the mass of all the particles within  concentric spheres, centred in the AHF position corresponding to the highest density peak, up to $r=2R_{200}$. We have interpolated the total mass profiles at fixed overdensities. 
We selected 24 linearly spaced overdensities, $\Delta$ from 200 to 2500, to homogeneously sample the profiles. This profile sampling allows us to predict the mass at overdensities that are commonly used in literature, such as $M_{200}$, $M_{500}$ and $M_{2500}$. With this approach, we differ and extend the previous work in literature that estimates the cluster mass always at a specific single aperture. $M_{200}$ was derived in \cite{Gupta2020} from simulated SZ map, \cite{Yan2020} by using optical, X-ray and SZ images generated from BAHAMAS simulation while \cite{Ntampaka2019} trained a CNN model to recover $M_{500}$ from Illustris TNG X-ray mock images.


The SZ \citep{SZ1970} effect consists of an inverse Compton scattering of cosmic microwave background (CMB) photons on the hot plasma of the ICM. This leaves a specific fingerprint on the CMB at the position of a galaxy cluster, shifting photons energy to higher frequencies. 
The observable of the SZ effect is the Compton-$y$ parameter defined as:

\begin{equation}
    y=\frac{\sigma_\mathrm{T} k_\mathrm{B}}{m_\mathrm{e} c^2} \int{n_\mathrm{e} T_\mathrm{e} dl},
	\label{eq:y_int}
\end{equation}

where $\sigma_T$, $k_B$, $c$ and $m_e$ are the Thomson cross-section, the Boltzmann constant, the speed of light and the electron mass at rest, respectively. Whereas, $n_e$, the electron number density, and $T_e$, the electron temperature, are integrated along the line of sight, $dl$.

In numerical simulation the quantity $n_\mathrm{e}$ can be substituted with the discrete number of electrons in the gas particle $N_\mathrm{e}$, by assuming that $n_\mathrm{e} = N_\mathrm{e}/dA/dl$, where $dA$ is the projected area. Thus, the integral in the equation~(\ref{eq:y_int}) can be replaced by the sum \citep{Sembolini2013, LeBrun2015}:

\begin{equation}
    y \simeq \frac{\sigma_\mathrm{T} k_\mathrm{B}}{m_\mathrm{e} c^2 dA} \sum_i{N_\mathrm{e,i} T_\mathrm{e,i} W(r, hi)},
	\label{eq:y_sum}
\end{equation}

where $W(r, h_i)$ is the projected Smoothed Particle Hydrodynamical (SPH) kernel adopted in simulation with the smoothing length $h_i$ and used to spread each mass particle to the surrounding area.
This equation is implemented in the public package \textsc{PYMSZ} \citep{Cui2018, Baldi2018} that is used here to generate Compton parameter $y$-maps. For each map, gas particles inside a cube of side $2R_{200}$, centred in the cluster centre identified by AHF, are taken into consideration. The pixel size of each map is parameterized in terms of $R_{200}$, $i.e.$ a $R_{200}xR_{200}$ map is sampled with 128x128 pixels.  
In order to increase the statistics, for each cluster we produced 29 maps at different projections rotating the cluster around its centre. 

\section{The Proposed Method}
\label{sec:method}

\begin{figure*}
	\includegraphics[width=\textwidth]{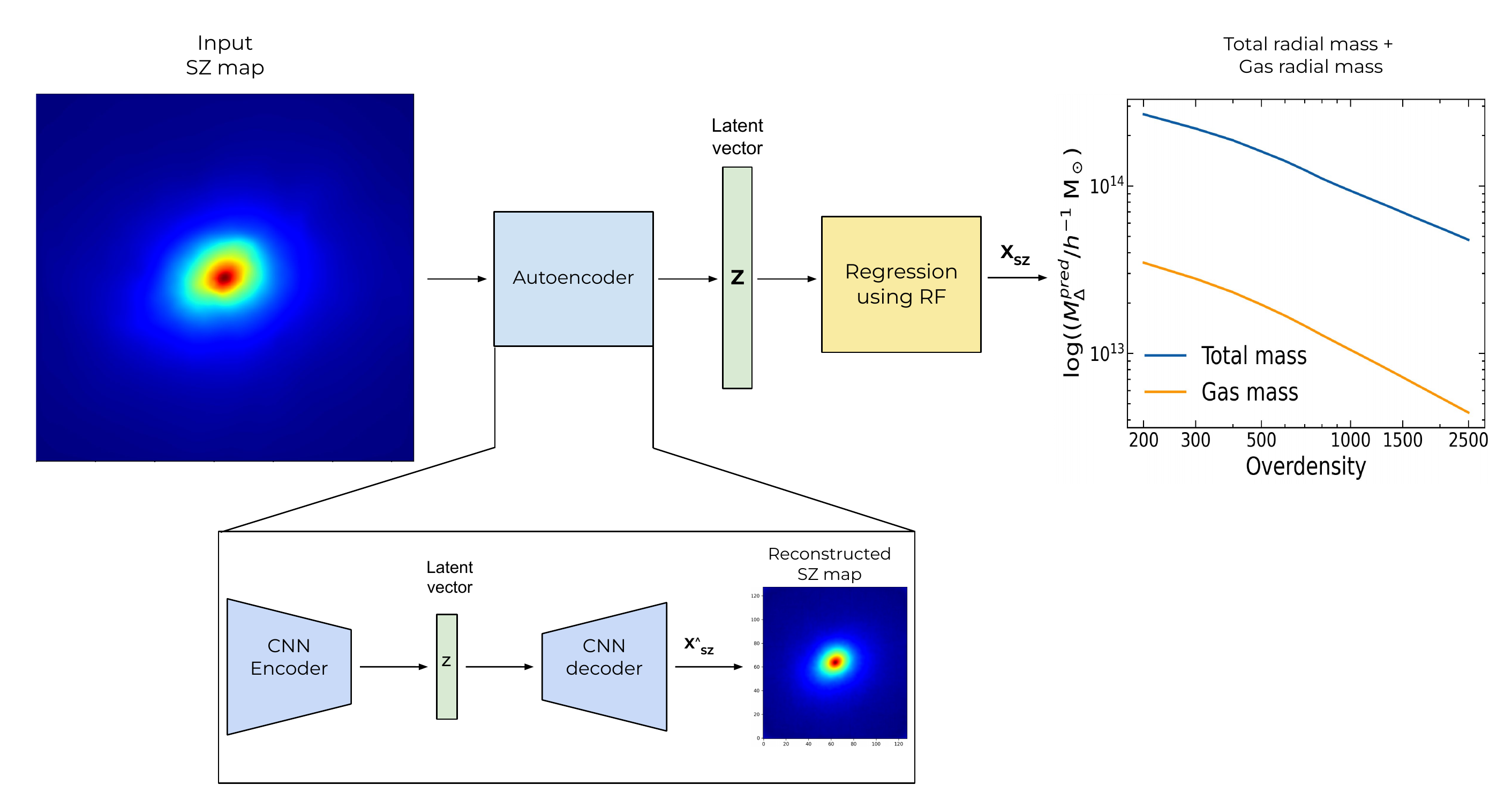}    
	\caption{The 3D radial mass profile inference architecture using the autoencoder plus the random forest. The autoencoder encrypts the input SZ map in a latent vector. The latter is then used as input for the RF regressor.
	}
    \label{fig:ae_arch}
\end{figure*}

The end-to-end pipeline is composed of two main parts: an autoencoder and a random forest regressor. The idea is to extract the features from the SZ images in an unsupervised manner and then to feed the obtained representation to an algorithm which learns how to predict the mass profiles. 
More details about the architecture can be found hereafter in the text, while the high level architecture is shown in Fig.~\ref{fig:ae_arch}.

\subsection{autoencoder for dimensionality reduction}

 An autoencoder is a neural network that tries to define a mapping between input $\mathbf{X}$ and an output (reconstruction) $\mathbf{\hat{X}}$ through an internal representation $\mathbf{Z}$, that has a dimension, $d$, smaller than the input one, $d_{in}$,~\citep{GoodBengCour16} and is used to learn useful properties of the data in an unsupervised setting. 
 It is composed of two parts: the encoder (mapping $\mathbf{X}$ to $\mathbf{Z}$: $\mathbf{Z}=f(\mathbf{X})$) and the decoder (mapping $\mathbf{Z}$ to $\mathbf{\hat{X}}$: $\mathbf{\hat{X}}=g(\mathbf{Z}$)). Recently, autoencoders have been used in astrophysics for different purposes, e.g. generative method of mock SZ observations \citep{Rothschild2022} and automatic morphological classification of galaxies \citep{Zhou2022}.
 
 In our contest, we set an autoencoder to derive a representative feature vector with a reduced dimension of our input data, i.e. the SZ maps, while being faithful to the original input. In particular, we build the encoder and decoder steps as follows:
 
\begin{itemize}
\item \textbf{Encoder} $\{E\}$: $X^{d_{in}} \rightarrow{Z^d}$:
it projects the input map $\mathbf{X_{SZ}}$ of dimension $d_{in}$ to a corresponding space through many stacked convolution-batch-normalization-ReLu layers. The output of each encoder is a $d$-dimensional vector $\mathbf{Z}$ that will be further used by the downstream task algorithm to infer the mass profiles. We note that $d \leq d_{in}$ since the encoder's role is the dimensionality reduction. In our architecture, the encoder is constituted by 4 layers that project the $d^{in}=128\times128$ SZ maps into a latent vector of dimension $d=150$.\\
\item\textbf{Decoder} $\{D\}$: $Z^{d} \rightarrow{X^{d_{in}}}$: it has a mirrored architecture to the encoder $E$ (4 layers). It takes a vector from the latent space of dimension $d=150$ and generates the corresponding map $\mathbf{\hat{X}_{SZ}}$ of dimension $d_{in}=128\times128$. Both networks $E$ and $D$ are trained using a reconstruction loss that catches the difference between the reconstructed map $\mathbf{\hat{X}_{SZ}}$ and the original one $\mathbf{X_{SZ}}$.
\end{itemize}
Thus, with this architecture, it acts as a self-supervised feature extractor.
For coding the autoencoder architecture, we made used of the publicly available \textsc{PyTorch}\footnote{\url{https://pytorch.org}} package \citep{pytorch2017}.

\subsection{random forest Method}
\label{sec:random_forest}
A random forest \citep[RF,][]{Breiman2001} is a supervised learning algorithm composed by a collection of decision trees. Each tree is an algorithm that is capable of performing classifications or regressions by entering as input some features and by applying a series of if-then-else statements until the possible conditions are fulfilled.
Although decision trees are a powerful tool, over time they have proven to be not very flexible 
and prone to overfitting. The combination of several trees in an RF overcomes these problems \citep{Segal2003}. This is achieved by assigning to each tree a subset of original data by bootstrap sampling and then all the predictions of the individual tree are averaged in the final result. This technique, known as "bagging", makes RF a robust and versatile model with low variance and less overfitting.

In this work, we use the class \textsc{RandomForestRegressor} implemented in the Python package \textsc{scikit-learn} \citep{pedregosa2011scikit}. In the RF set-up phase, we setup the hyper-parameters of the function in order to optimise the performance of the model. According to \cite{fernandez_14_do} and \cite{, bentejac_2020_comparative}, we identify the number of trees, \textsc{n\_estimators}, as the most important hyper-parameter, and the maximum depth of each tree, \textsc{max\_depth}, as the second. 
We observe that we reach convergence for $\textsc{n\_estimators} \geq200$ regardless of the depth of the trees. Regarding the depth, we obtain the best results with the default value of the \textsc{max\_depth} parameter. Consequently, we set the RF with 200 trees and all other hyper-parameters to default values of the \textsc{RandomForestRegressor} function.

\subsection{Train and Test sets Split}
\label{subsec:train_test}
Once the RF has been planted, we train it to predict the radial total and gas-only mass profiles from
the information extracted from the SZ maps by the autoencoder algorithm. Therefore, we perform a random selection of the training and test sets containing $80\%$ and $20\%$ of the original sample of 2522 clusters, respectively.
Subsequently, we increase the statistics of the two sets by taking into account the 29 projections for each cluster. We have thus ensured that the same cluster cannot belong to different datasets at the same time. Although the mass profiles are extracted from three-dimensional distributions and are common to each projection
in both the training and test phases, each projection was treated as independent. This was possible because the starting features are the information extracted from the autoencoder that are different for each map and for each projection.

\section{results}
\label{sec:results}

\begin{figure}
\centering
	\includegraphics[width=\columnwidth]{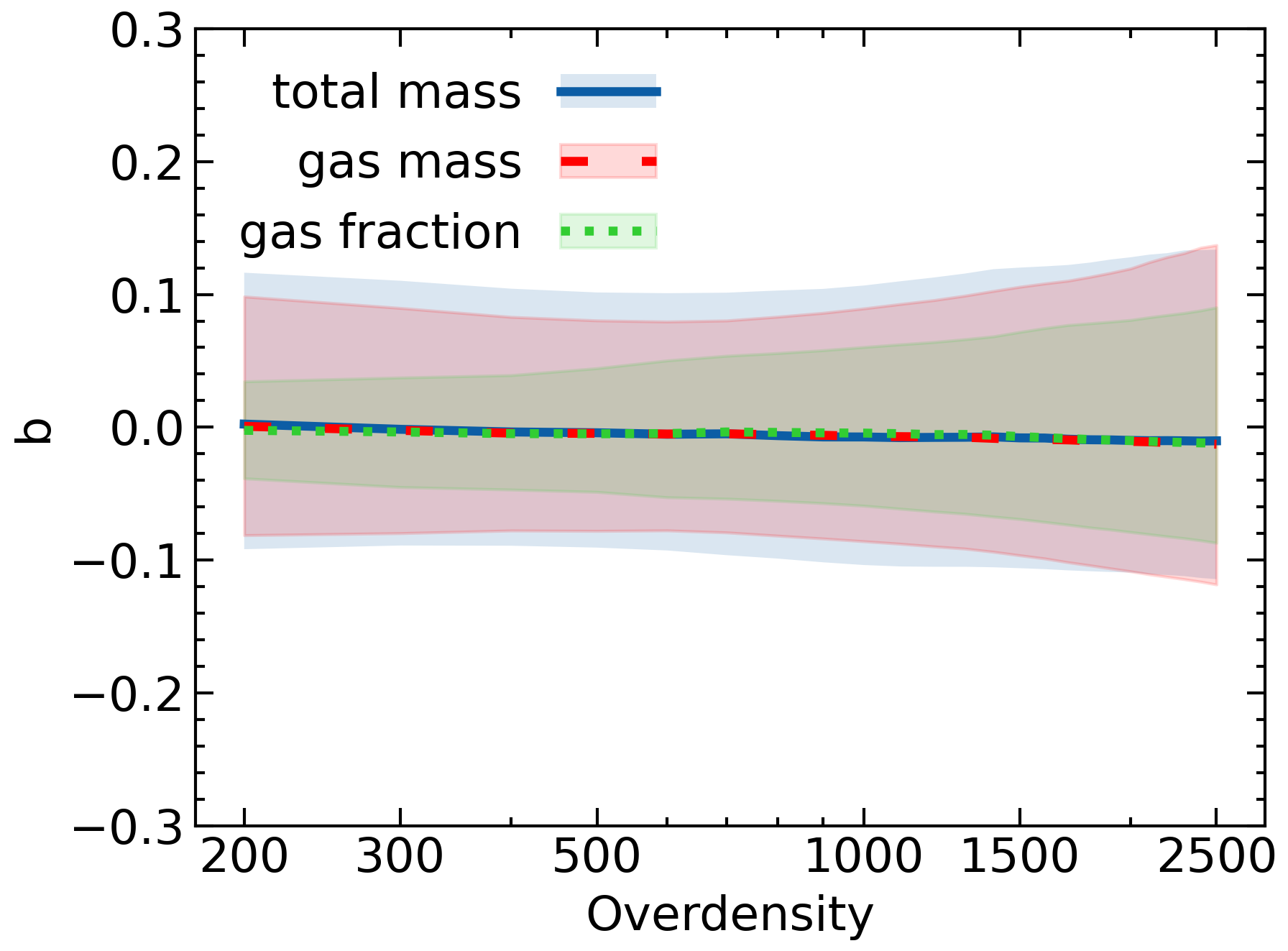}
    \caption{The median bias as a function of the overdensity for total mass (blue solid line), gas mass (red dashed line) and gas fraction (green dotted line). The shaded light blue, pink and green regions correspond to the $16^\text{th}-84^\text{th}$ percentiles for total, gas mass and gas fraction, respectively.}
    \label{fig:bias_overdensity}
\end{figure}

In this section, we present the application of the trained RF on the test set and the accuracy of its predictions.

We analyse the performance of our model by comparing the predicted and the true profiles at each overdensity. 
We analyse the performance of our model by comparing the predicted profiles and the true one of the test set, at each overdensity. To perform this task, we define the bias at each $\Delta$ as:
\begin{equation}
    b_{\Delta} = \frac{M_{\Delta}^\text{pred}-M_{\Delta}^\text{true}}{M_{\Delta}^\text{true}}.
	\label{eq:bias}
\end{equation}

The biases in the cluster mass estimate after the training of our algorithm is shown in Fig.~\ref{fig:bias_overdensity}, where the median bias $b$ is evaluated as a function of the overdensities
for total and gas masses, represented with blue solid line and red dashed line, respectively. The masses predicted by the RF model are unbiased in the whole range of overdensities, from the cluster core to the outskirts. The median values increase towards the centre but always less than $\sim1\%$.
The scatter, quantified with the $16^\text{th}$ and the $84^\text{th}$ percentiles and shown with the shaded regions, is around $10\%$ for the total mass (blue) and $\sim 8\%$ for the gas mass (red). Moreover, the scatter is not properly constant, but become slightly larger in the direction of the cluster outskirts and towards the cluster centre. The minimum scatter is reached for overdensities around 500 and 600 for total and gas profiles, respectively.

The bias for the gas fraction, defined as $(f_\text{gas}^\text{pred}-f_\text{gas}^\text{true})/f_\text{gas}^{\text{true}}$, is less than $1\%$ in the entire overdensities range considered in this analysis. In this case the scatter decreases from $\sim10\%$ to $\sim 3\%$, as shown with the green dotted line in Fig.~\ref{fig:bias_overdensity}. 

\subsection{Bias dependence on the total cluster mass}
\label{subsec:mass_dep}
\begin{figure}
	\includegraphics[width=\columnwidth]{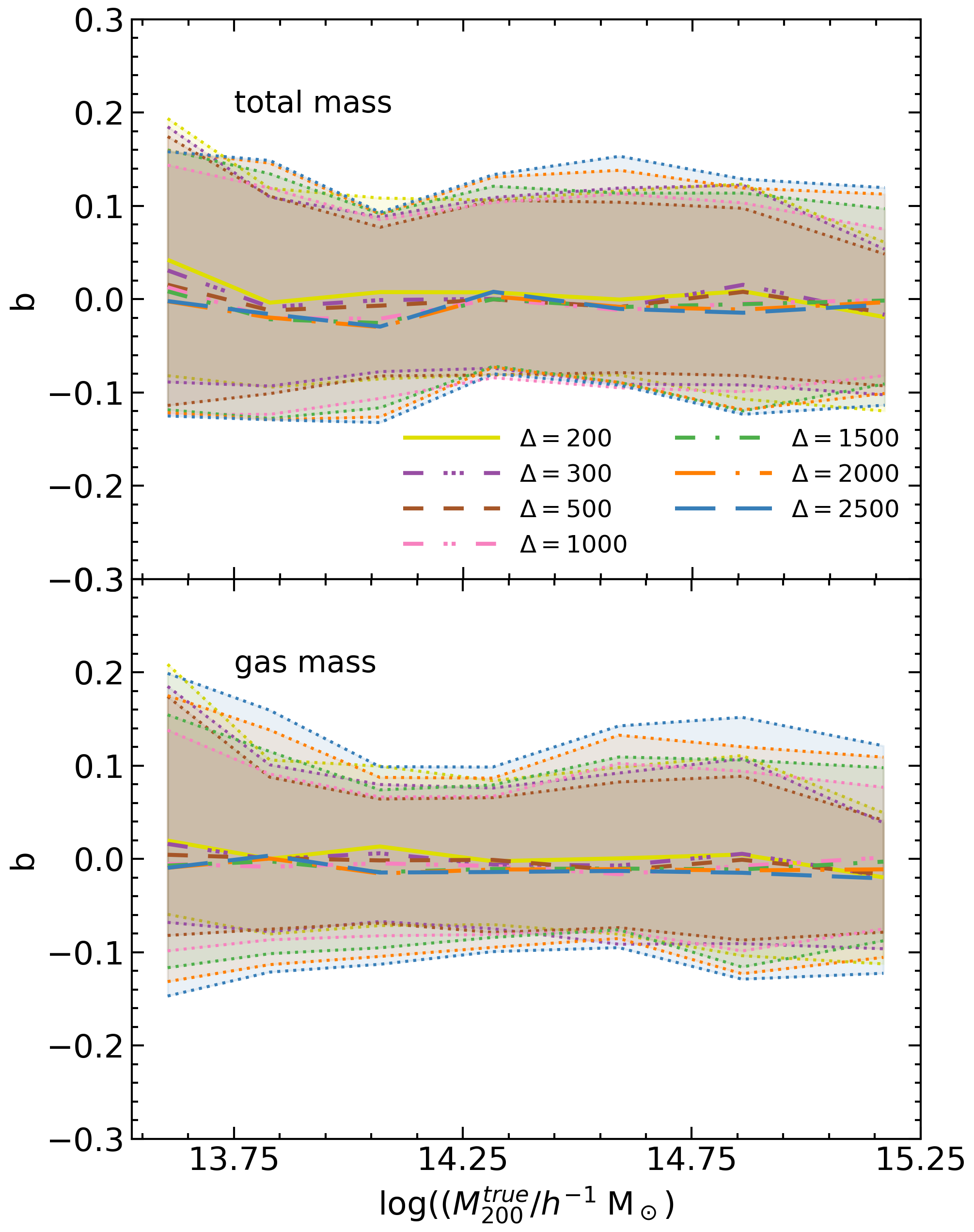}
	
    \caption{The median bias of the predicted total (upper panel) and gas (lower panel) as a function of the logarithm of true mass $M_{200}^\text{true}$. For graphical reasons we only show the lines representing the bias at $\Delta=200, 300, 500, 1000, 1500, 2000, 2500$ in yellow, purple, brown, pink, green, orange and blue, respectively. The shaded areas correspond to the $16^\text{th}-84^\text{th}$ percentiles intervals.}
    \label{fig:dep_mass}
\end{figure}

To test if the performance of our ML model suffers of any particular bias related to the true cluster total mass, $M_{200}^\text{true}$, we study the mass dependence of ML predictions dividing the sample in seven equally populated mass bins. We show in Fig.~\ref{fig:dep_mass} the median and relative scatter of the total (upper panel) and gas (lower panel) biases for seven chosen overdensities (for graphical reasons). In general, we do not observe any dependency of gas mass profile on cluster true masses in the whole range of  overdensities considered. Whereas, we see a tendency to overestimate the total mass in the first mass bin ($10^{13.5}\leq M_{200}^\text{true}/\hMsun< 10^{13.7}$) for overdensity $\Delta = 200$ and $300$. The scatter is around $10\%$ but tends to increase in the low mass bins for both total and gas profiles. 

\subsection{Bias dependence on dynamical state}
\label{subsec:chi_dep}
\begin{figure}
	\includegraphics[width=\columnwidth]{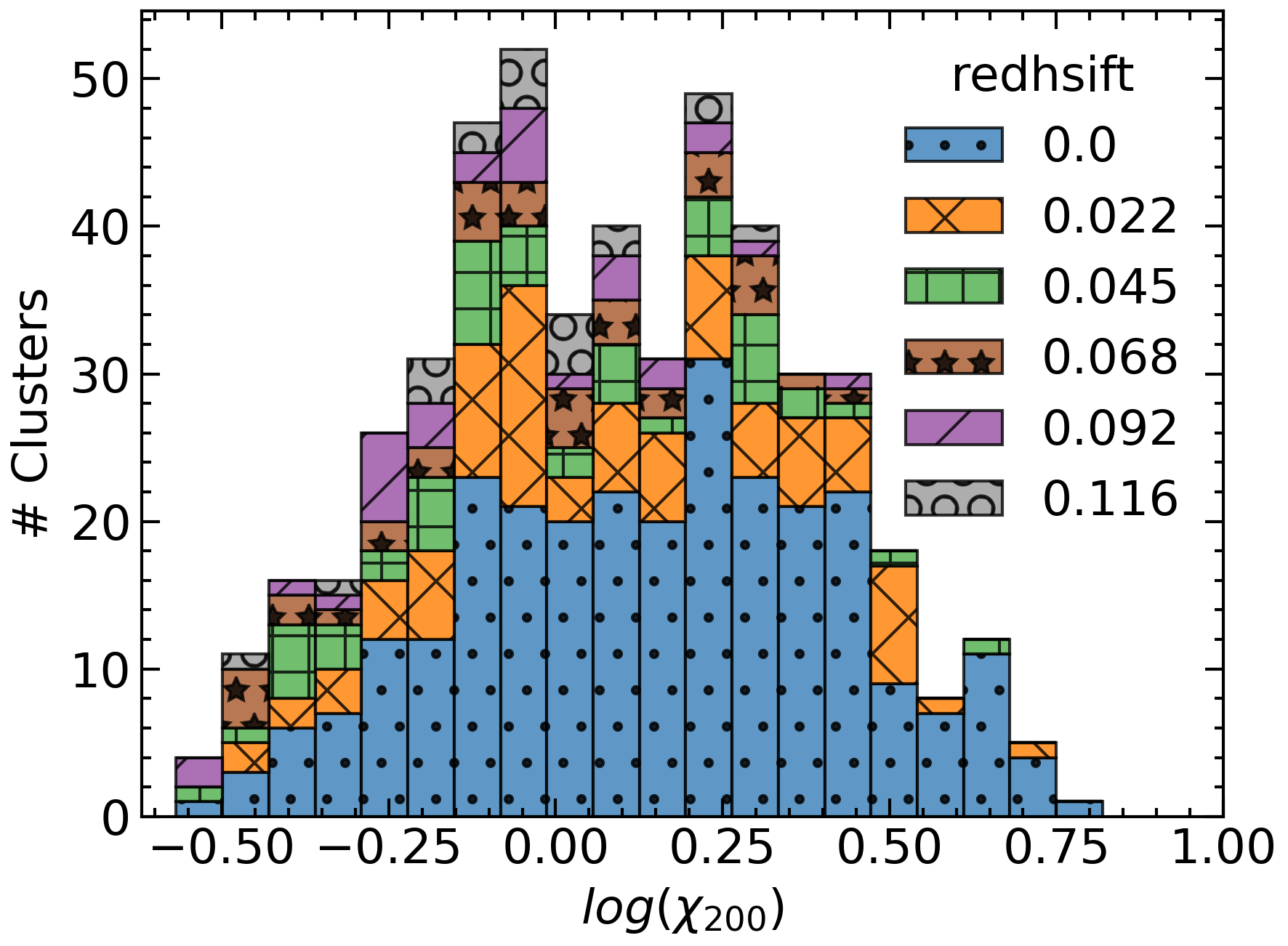}
    \caption{Classification of the dynamical state of the clusters within the test sample in terms of the relaxation parameter $\chi_{200}$. The blue bars represent zero redshift clusters, while higher redshift clusters are represented by the coloured bars according to the legend.} 
    \label{fig:distribution_chi}
\end{figure}
Classical methods to infer the cluster mass from SZ and X-ray observations make assumptions on the hydrostatic equilibrium of the clusters, so they are sensitive to deviation from it \citep[e.g. ][]{Ruppin2018, Pearce2020, Gianfagna2021}. Therefore, it is extremely important to investigate if and how the performance of our ML approach change with the dynamical state of the clusters. 
Considering hydrodynamical simulation, it is possible for a specific cluster to extract any possible dynamical or thermodynamical information about its components, like particle 3D velocity, entropy, etc. Based on these information, several indicators have been defined in the literature to assess the dynamical state of synthetic clusters. In this work,
we use the relaxation parameter $\chi$ originally  introduced in \cite{Haggar2020} and later revised in \cite{DeLuca2021}, combining only 2 indicators:

\begin{equation}
    \chi_{200} = \sqrt{\frac{2}{\left(\frac{f_s}{0.1}\right)^2+\left(\frac{\Delta_r}{0.1}\right)^2}}.
	\label{eq:chi_par}
\end{equation}
where $f_s$ is the ratio between the sum of the masses of all the sub-halos within $R_{200}$ and the cluster total mass $M_{200}$, and $\Delta_r$ is the offset between the theoretical centre of the cluster and the centre of mass of the cluster, normalised to $R_{200}$. The distribution of our sample as a function of the relaxation parameter is shown in Fig.~\ref{fig:distribution_chi}, here negative and positive tails represent extremely disturbed and extremely relaxed systems, respectively.

\begin{figure}
	\includegraphics[width=\columnwidth]{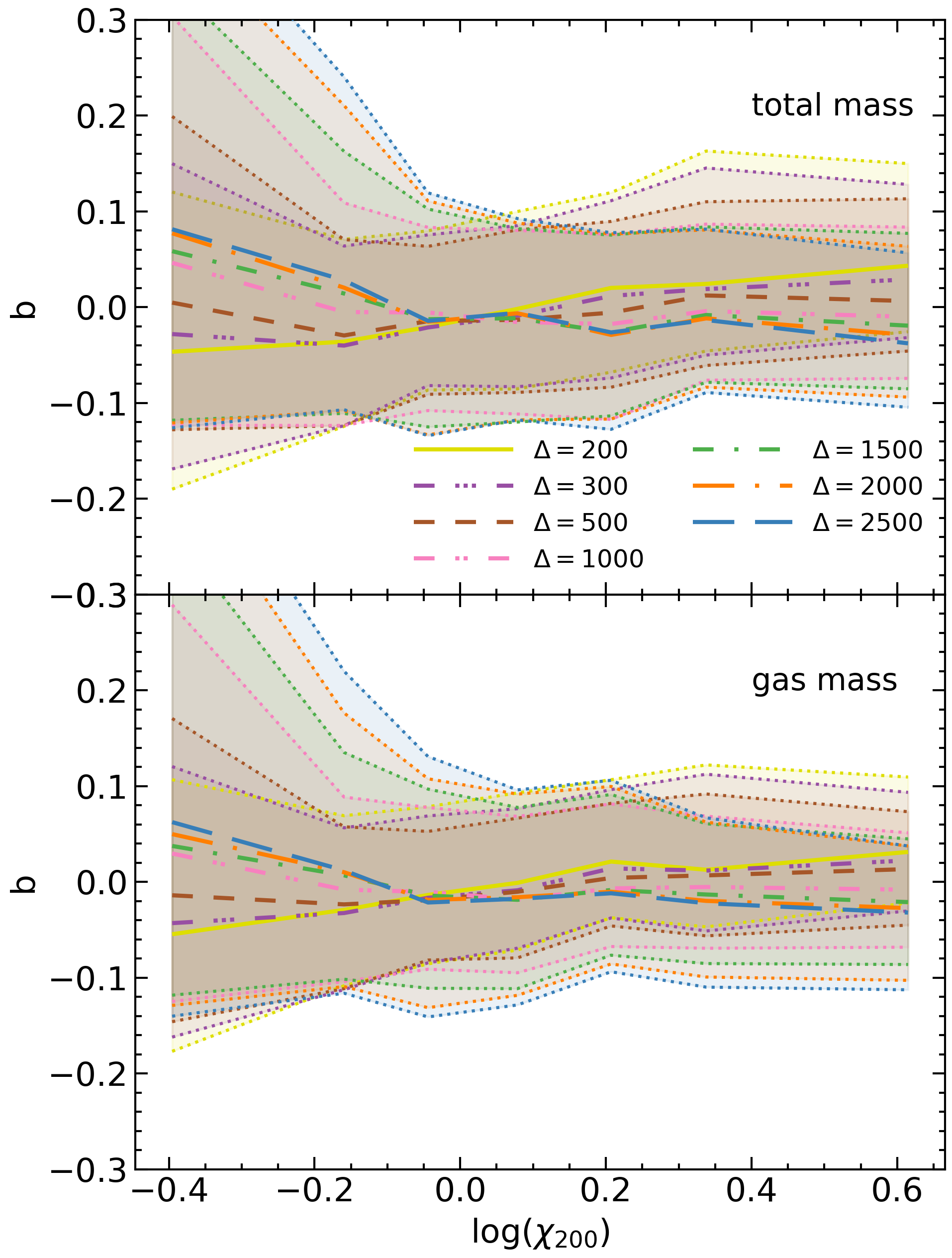}
    \caption{The median bias of the predicted total (upper panel) and gas (lower panel) mass with respect $M_\text{true}$ as a function of the logarithm of the relaxation parameter $\chi_{200}$. For graphical reasons we only show the lines representing the bias at $\Delta=200, 300, 500, 1000, 1500, 2000,
    2500$ in yellow, purple, brown, pink, green, orange and blue, respectively. The shaded areas correspond to the $16^\text{th}-84^\text{th}$ percentiles intervals.}
    \label{fig:dep_chi}
\end{figure}

The median biases are shown in Fig.~\ref{fig:dep_chi} for seven equally populated bins of $\chi_{200}$. Only seven overdensities are plotted for clarity. We see that in general, the model is sensitive to the dynamical state of the clusters.
There is a dependence of both total (upper panel) and gas (lower panel) mass reconstruction with the dynamical state, depending also on the overdensity. Our ML analysis tends to underestimate the mass in the outskirts of disturbed clusters and to overestimate it in the inner part, 
while it behaves in the opposite for relaxed systems. However, profiles at $\Delta=500$ and $600$ do not show any dependence  on $\chi_{200}$. This behaviour, as a function of the dynamical state, can also explain the minimum of the scatter around these overdensities that we observed in Fig.~\ref{fig:bias_overdensity}.
Regarding the scatter, it remains $\sim10\%$ for $\log(\chi) > 0$ at all the overdensities. For $\log(\chi)<0$, the scatter starts to grow as the clusters become more disturbed. 
Moreover, we see that the scatter remains around $15\%$ for overdensities related to the cluster outskirts, whereas it grows up to $\sim30\%$ going towards clusters' inner regions.

\section{The Concentration- Mass relation from  inferred profiles}
\begin{figure}
   \centering
   \includegraphics[width=\columnwidth]{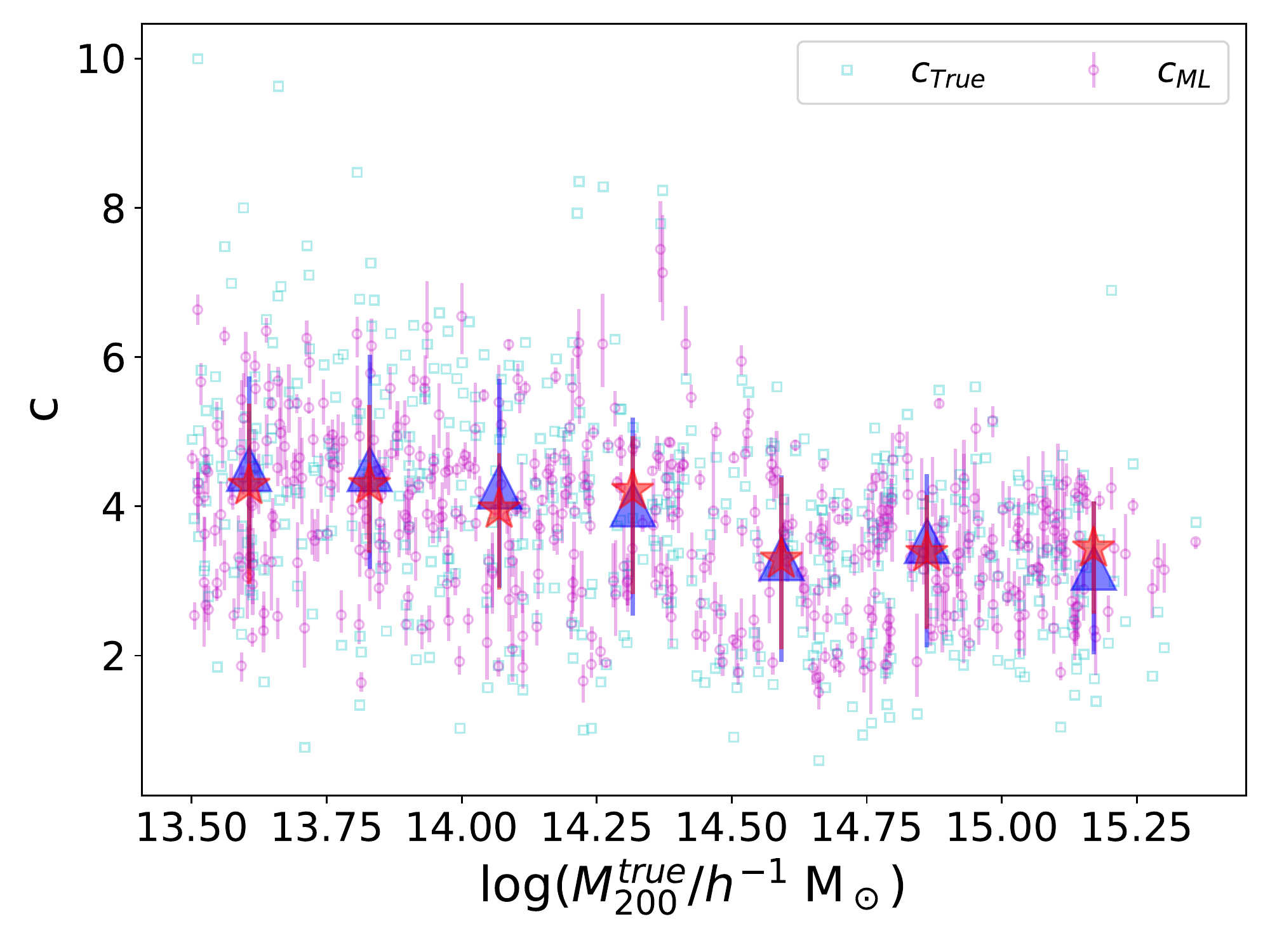}
   \caption{The concentration - halo mass  relation from both the true profiles, denoted  by $c_\text{true}$, and the ML profiles $c_\text{ML}$. The blue triangles and red stars show the median concentrations for $c_\text{true}$ and $c_\text{ML}$ respectively within the halo mass bins. Error bars show the $16^\text{th}-84^\text{th}$ percentiles. Magenta circles with error bars to show the median value with $16^\text{th}-84^\text{th}$ percentiles of the 29 projections. 
   }
   \label{fig:c_mas1}
\end{figure}

\begin{figure}
    \centering
    \includegraphics[width=\columnwidth]{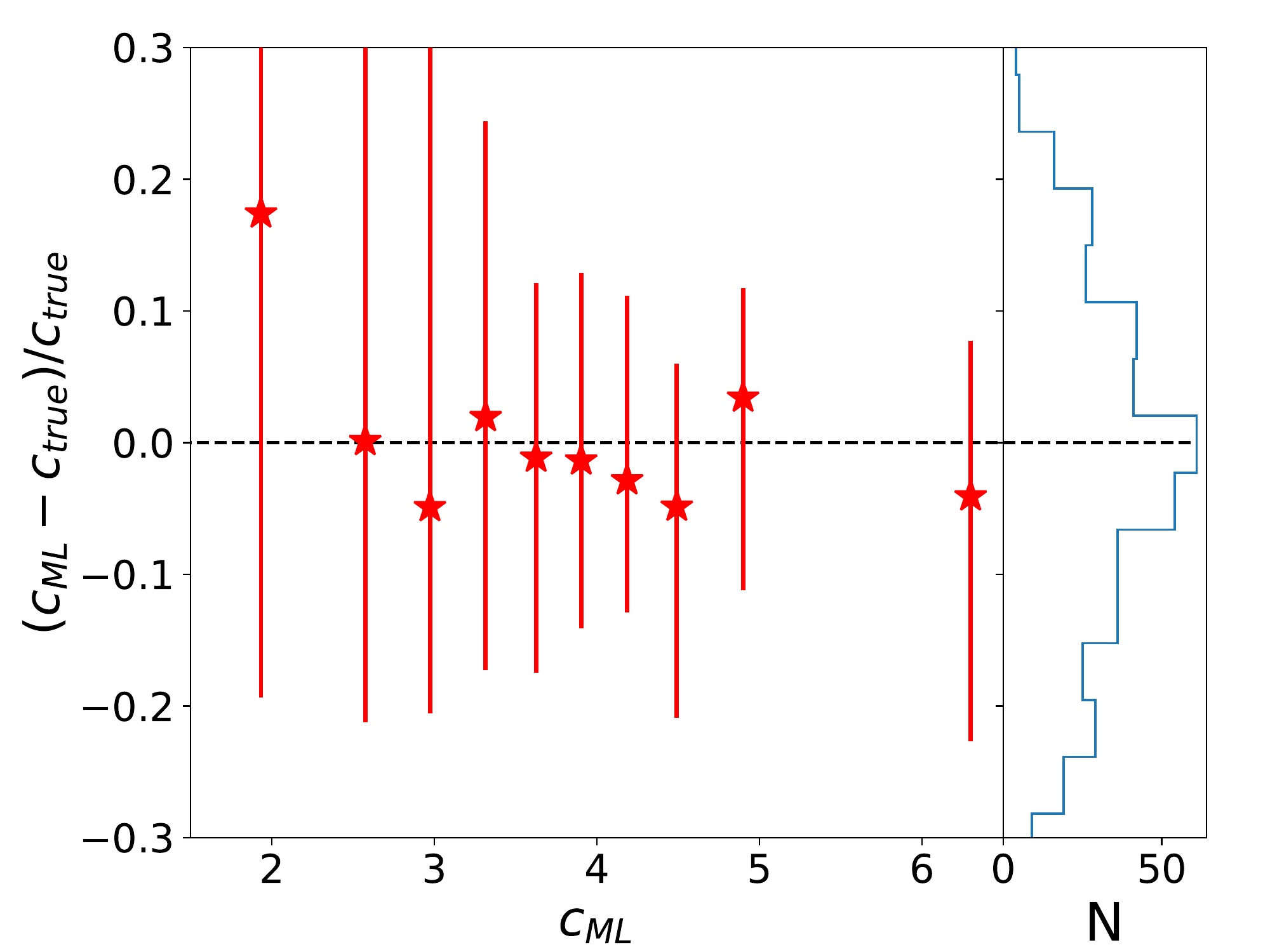}
    \caption{The concentration bias as a function of $c_\text{ML}$ from the predicted halo profile. Here we only show the median values and $16^\text{th}-84^\text{th}$ percentiles in each concentration $c_\text{ML}$ bin which contains the same number of clusters per bin. Thus a slight difference in bin size is expected. Here, $c_\text{ML}$ means the median result of concentration from the 29 projected maps  per cluster. The right-hand side panel shows the overall distribution of the concentration bias. 
    }
    \label{fig:c_mas2}
\end{figure}

Investigating the mass profile provides more information on the galaxy cluster internal structure.  
A general, good tracer of the mass profile of galaxy clusters is given by the two-parameter Navarro-Frenk-and-White (NFW) model \citep{Navarro1997}. Besides the enclosed masses estimated at different overdensities, we examine whether the profiles from these masses give consistent NFW concentration parameters.
In this section, we fit the predicted profiles with 24 different data points (overdensities) to the NFW profile and derive the $c_\text{ML} = R_{200}/r_s$, where $r_s$ is the typical scaling radius of the NFW profile. 
Thus, we compare these estimates with the concentration parameters, $c_\text{true}$, calculated using the true mass profiles.

In Fig.~\ref{fig:c_mas1}, we show both $c_\text{true}$ (cyan squares) and $c_\text{ML}$ (magenta cycles) as a function of the true halo mass $M_{200}^\text{true}$. The errorbars for the $c_\text{ML}$ data (magenta circles) represent $16^\text{th}-84^\text{th}$ percentiles among the 29 projections.
The concentration-mass relation, represented by the blue triangles and red stars, is obtained by computing the median of $c_\text{true}$ and $c_\text{ML}$ in seven equally populated mass bins, respectively. The relations of our ML analysis and the true one are almost indistinguishable, whereas the errorbars ($16^\text{th}-84^\text{th}$ percentiles) in the ML $c-M$ relation are slightly smaller than the true ones.
In general, the predicted concentration has small scatters with respect to the true one, which means a small projection effect for these $c_\text{ML}$. Therefore, the extreme concentration values at both $c \lesssim 1$ and $c \gtrsim 7$ are less represented. 

We further quantify the concentration bias (${c_\text{ML}}/c_\text{true} -1$) in Fig.~\ref{fig:c_mas2} as a function of the predicted median concentration, ${c_\text{ML}}$, over the 29 projections.
 
After dividing the sample into ten bins of ${c_\text{ML}}$ with the same number of clusters, the median concentration bias wiggles around 0, except at the lowest concentration bin. However, it is worth noting that the error bars at low concentrations are larger compared to those of mass biases (Fig.~\ref{fig:dep_mass}), while they are comparable at intermediate and high concentrations. Instead, we see that this trend is similar to the one shown in Fig.~\ref{fig:dep_chi}. We speculate that a possible explanation for this behaviour is that ML tends to overpredict the masses at higher overdensities (see Fig.~\ref{fig:dep_chi}) for dynamical unrelaxed clusters caused by major merger events that, instead, tend to have lower concentrations. Therefore, the concentration of ML profiles is biased high for objects with low concentrations.
concentrations, while ML underpredict their masses in the core, which results in a slightly lower concentration. This picture is consistent with having fewer outliers from the ML results, as shown in Fig.~\ref{fig:c_mas1}.

\section{Comparison with HE mass estimates}
\label{sec:discussion}
\begin{figure}
\centering
	\includegraphics[width=\columnwidth]{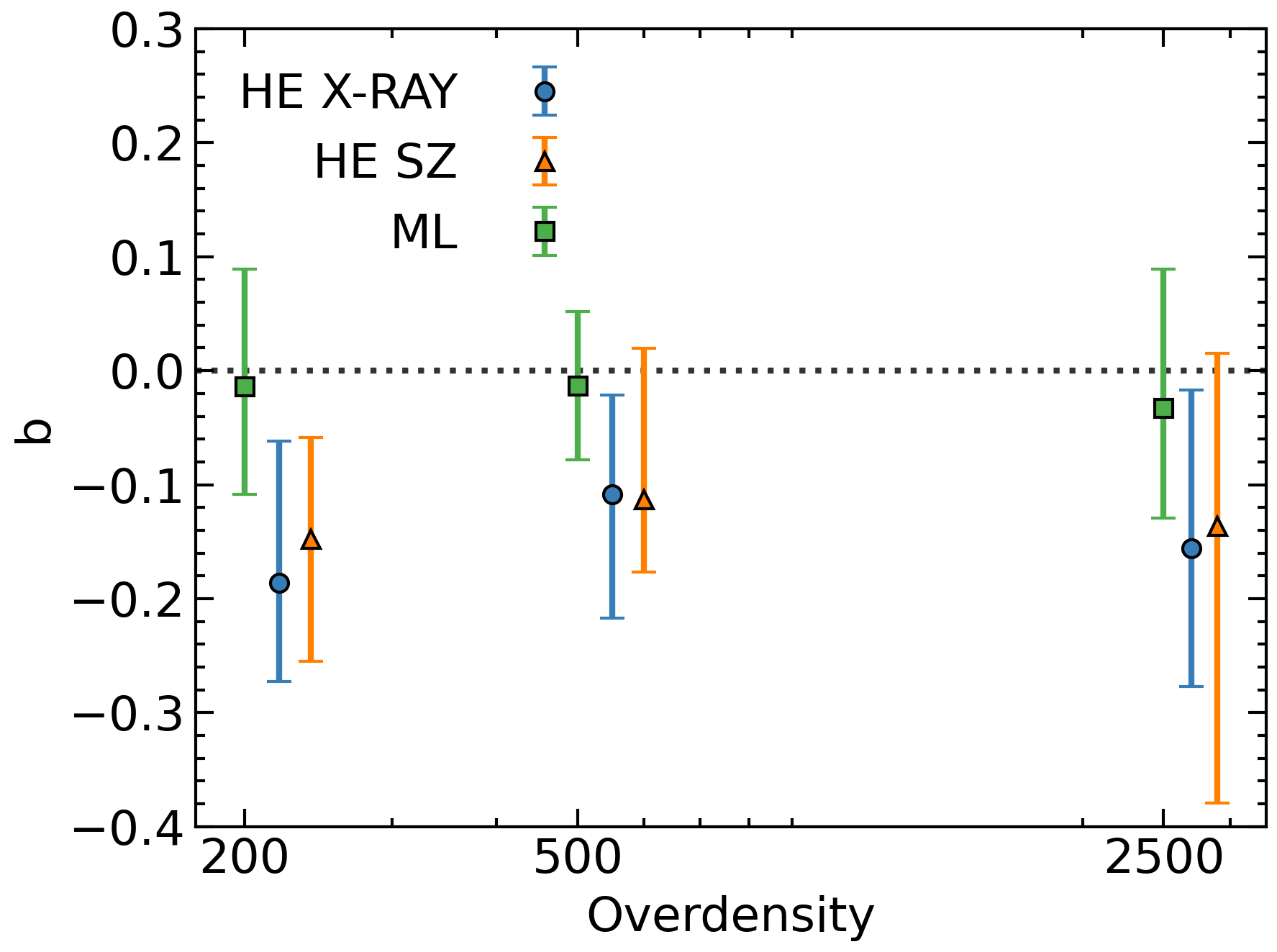}
    \caption{Median bias estimated with ML method compared with HE mass bias values at the 3 common overdensities for the same clusters selection in \thethreehundred{} sample at $z=0$ analysed in \citet{Gianfagna2022}. The error bars represent $16^\text{th}-84^\text{th}$ percentiles.}
    \label{fig:Giulia_comparisons}
\end{figure}
Our ML model recovers the  mass radial profiles with median bias close to zero. It is remarkable that we achieve this result without making any a priori assumption on the physical properties of the clusters. 
Here we compare our median bias with the  mass bias computed  when HE approximation is adopted. We refer to the analysis on \cite{Gianfagna2022} where also synthetic clusters from \thethreehundred{} have been used. The mass of the clusters was inferred by using ICM data typical of X-ray and SZ observations. Only the most massive clusters present in each resimulated region at redshift $z$=0 were considered. In order to compare our result with them, we recalculated the bias only for the common clusters, 53 objects in a mass range between $1.3\times 10^{14}\, \hMsun$ to $3\times 10^{15}\, \hMsun$.
The results of this comparison are shown in Fig.~\ref{fig:Giulia_comparisons}. The HE bias is of the order of $10-20\%$ considering X-ray observables (red dots) such as electron gas temperature and density, or SZ (green) derived pressure and density.
In the case of ML estimates (blue dots), it is clear that the bias is less than $1\%$ with a scatter around $10\%$. Interestingly, the bias shows the lower scatter at $\Delta =500$. Although the biases are compatible within the errors, ML mass estimates are systematically more accurate and unbiased than the HE ones. Moreover, the ML approach results in a smaller scatter in the whole range of overdensities.

\section{conclusions}
\label{sec:conlusions}
 
 Only recently cluster masses have been recovered with ML approaches starting from different spectral bands images but always at one specific single aperture, such as $M_{200}$ or $M_{500}$. In this paper, we present an ML model that for the first time is able to infer simultaneously the full integrated radial profile for the gas and the total mass from SZ mock images. 
The ML model architecture is a combination of an autoencoder and a random forest regressor. This ML algorithm is trained and tested on a sample of 73,138 mock Compton-$y$ parameter maps generated along 29 projections for each of the 2,522 galaxy clusters extracted from the \gadgetx{} run of the \thethreehundred{} simulations.

The autoencoder is used to encrypt the relevant information from each map, while the random forest performs the final estimation of the radial mass profiles. The model is able to infer the gas mass profile, responsible for the SZ effect signal in the maps, but also the cluster total mass without any apriori assumption on the hydrostatic equilibrium of the cluster.

 Our main results can be summarised as follows:
 
\begin{itemize}
    \item The ML model is able to recover unbiased profiles (bias lower than $1\%$) with a scatter of $\sim10\%$ that is slightly increasing towards the outskirts and to the inner part of the cluster (Fig~\ref{fig:bias_overdensity}) with a minimum of around an overdensity  of $\sim  600$;
    \item The accuracy and the precision of the method do not depend on the cluster mass (Fig~\ref{fig:dep_mass}). Nevertheless, they are more affected by the dynamical state with an impact that depends on the overdensity. In general, the scatter increases in unrelaxed clusters due to projection effects;
    \item From the total and gas mass profiles, we also derived the gas fraction profile. The bias, in this case, is also lower than $1\%$ while the scatter decreases up to $\sim 3\%$ in the outskirts;
    \item The concentration parameter, obtained by fitting the inferred total mass profiles with an NFW model, shows to be unbiased with a scatter between 10\% and 20\% for $c_\text{ML}$>2. Therefore, the ML predicted $c$-$M$ relation is in reasonable agreement with the true one. 
    \item The comparison with a standard method to infer the cluster total mass, such as the HE approximation, shows that our estimation of the mass is more accurate as it does not suffer from the hydrostatic mass bias.
    
\end{itemize}

Furthermore, in order to make this approach less prone to the physical models implemented in the simulation, we investigated the possibility of training the model with data from different simulations, such as \gadgetx{} (AGN feedback) and \gizmo{} (strong AGN feedback). As described in Appendix \ref{sec:appendix}, the network trained in this way is able to marginalise over different hydrodynamical simulations, obtaining results that are compatible with single-simulation training.  
Therefore, for future applications, we are going to use this kind of approach by further extending the data set with other simulations is of paramount importance in order to  marginalise over all possible baryonic effects.

The framework presented in this paper  can be extended  to infer also other ICM radial profiles such as gas temperature and pressure, among others. Its  application on different observational maps, such as optical and X-ray mock images,  seems a  promising way to improve the reconstruction of these  profiles.

In a forthcoming  paper, we plan to  train this ML approach on mock SZ maps, adding instrumental and observational effects, such as noise and limited angular resolution. Then, the model will be tested on real Compton-$y$ parameter maps at  different angular resolutions and at different redshifts. Nevertheless, the redshift evolution of the $Y-M$ relation is negligible or weak up to z=1 \citep{Henden2019,deandres2022baryon}. The bias in the inferred ML mass will  also be compared with other methods using weak lensing mock images of the same cluster dataset.

\section*{Acknowledgements}
We thank the anonymous referee for his/her valuable suggestions and comments.
Most of the simulations used in this work  have been performed in the MareNostrum Supercomputer at the Barcelona Supercomputing Center, thanks to CPU time granted by the Red Espa\~nola de Supercomputaci\'on.
AF, AS and MDP acknowledge support from Sapienza Università di Roma thanks to Progetti di Ricerca Medi 2019, RM11916B7540DD8D and 2021, RM12117A51D5269B. AF also thanks financial support by Universidad de La Laguna (ULL), NextGenerationEU/PRTR and Ministerio de Universidades (MIU) (UNI/511/2021) through grant "Margarita Salas".
DdA, WC and GY acknowledge Ministerio de  Ciencia e Innovación (Spain) for partial financial support under research grant PID2021-122603NB-C21.
WC is also supported by the STFC AGP Grant ST/V000594/1 and the Atracci\'{o}n de Talento Contract no. 2020-T1/TIC-19882 granted by the Comunidad de Madrid in Spain. He further acknowledges the science research grants from the China Manned Space Project with NO. CMS-CSST-2021-A01 and CMS-CSST-2021-B01.

\section*{Data Availability}

The data used in this paper is part of The Three Hundred Project and can be accessed following the guidelines  that can be found in the main website \footnote{\url{https://the300-project.org}} of the collaboration. The data specifically shown in this paper will be shared upon request to the authors.
 



\bibliographystyle{mnras}
\bibliography{Biblio.bib} 

\appendix
\section{Impact of multi-simulation training}
\label{sec:appendix}
In Sec.~\ref{sec:dataset} and \ref{sec:method} we described that the proposed network was trained with cluster maps selected within the \gadgetx{} run of \thethreehundred{} project simulation. While in Sec.~\ref{sec:results} we showed the quality of the results obtained on a test set composed of clusters selected from the same simulation. However, each simulation has its own characteristics such as cosmology, resolution or different baryonic physics. These differences can have effects of greater or lesser importance on e.g. the mass of the structures, the shape of the mass profiles or the SZ maps. For this reason, if the differences between the products of different simulations are significant, the accuracy of ML or more classical methods, such as scaling relations, may be compromised. In the context of machine learning, to overcome this problem one possible approach is to train a network with the products of different simulations as proposed by the Cosmology and Astrophysics with MachinE Learning Simulations \citep[CAMELS, ][]{Camels}. Following this approach, we decided to re-train our model by adding to the clusters selected from the \gadgetx{} run those produced by the \gizmo{} run, for details on the the two runs, see \citep{Cui2022}.

\begin{figure}
\centering
	\includegraphics[width=\columnwidth]{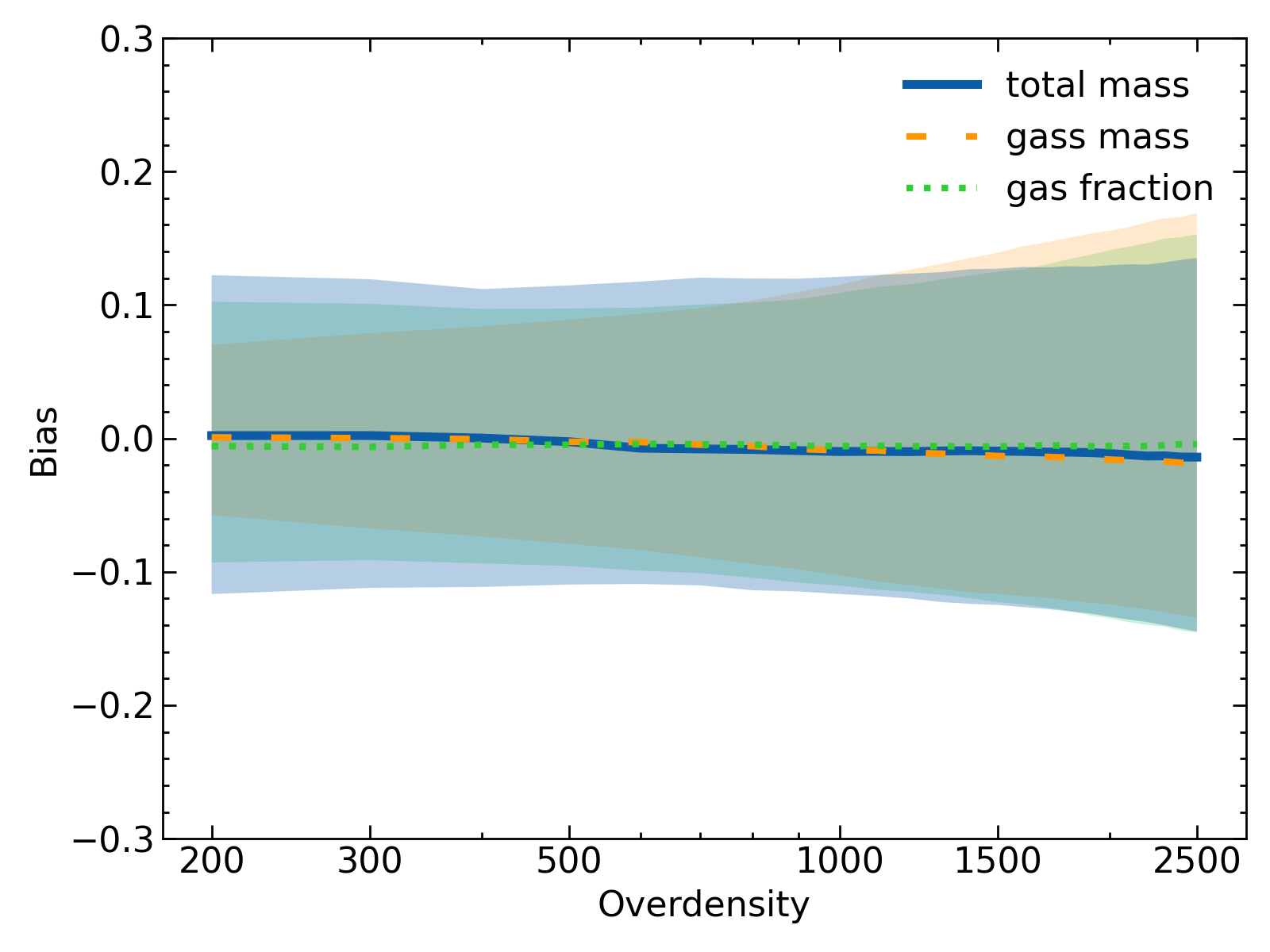}
    \caption{The median bias of a sample of clusters selected from both \gadgetx{} and \gizmo{} as a function of the overdensity for total mass (blue solid line), gas mass (orange dashed line) and gas fraction (green dotted line). The shaded light blue, orange and green regions correspond to the $16^\text{th}-84^\text{th}$ percentiles for total, gas mass and gas fraction, respectively. }
    \label{fig:bias_mix}
\end{figure}

Figure~\ref{fig:bias_mix} shows that using the network to predict the mass profiles of a test set consisting of a mixture of clusters from \gadgetx{} and \gizmo{} yields results that are completely in line with what was obtained with the network trained and tested on \gadgetx{} clusters alone, i.e. the median bias of both mass profiles is zero and the scatter is about 10\% for all overdensities.

\begin{figure}
\centering
	\includegraphics[width=\columnwidth]{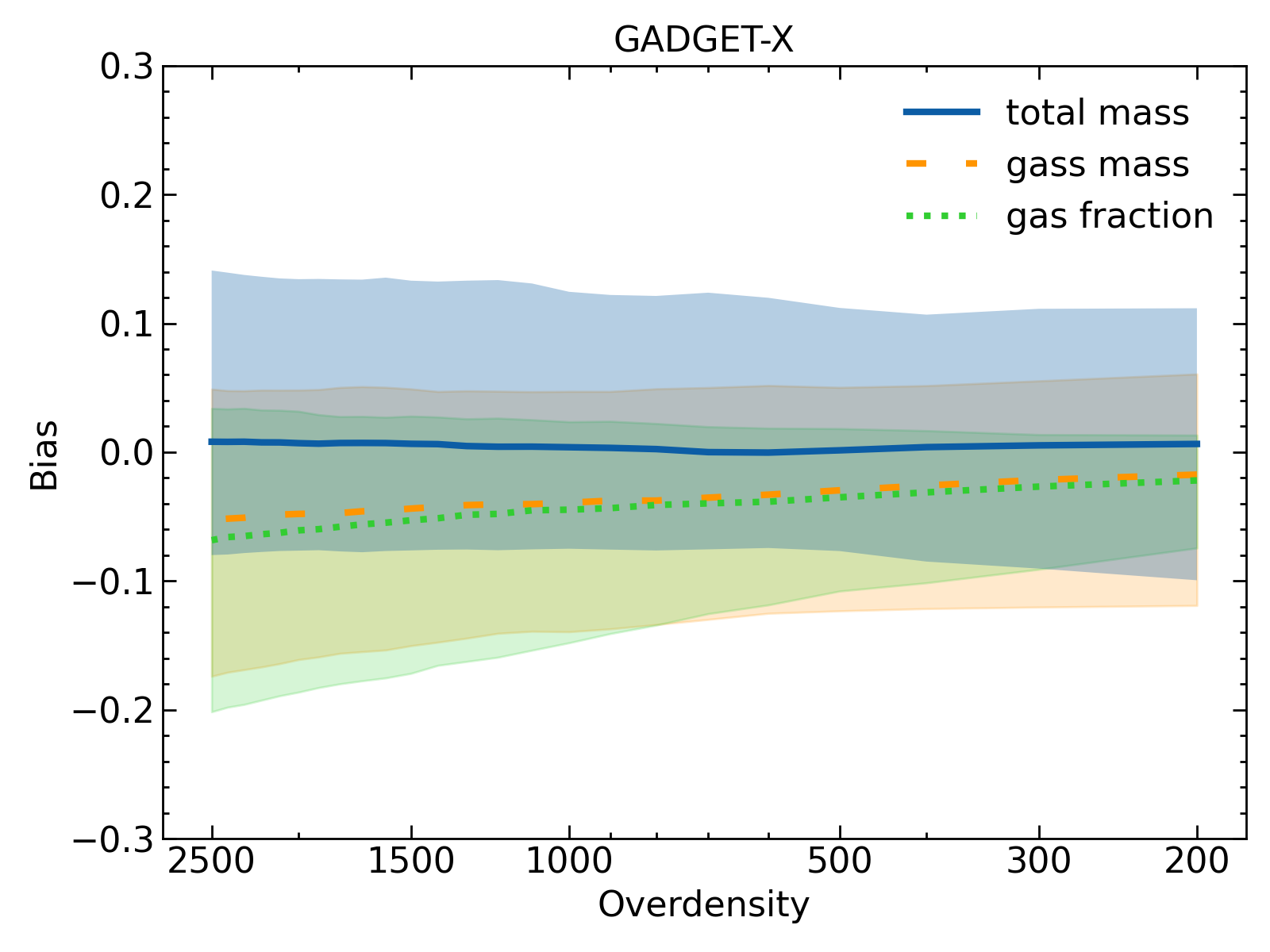}
    \includegraphics[width=\columnwidth]{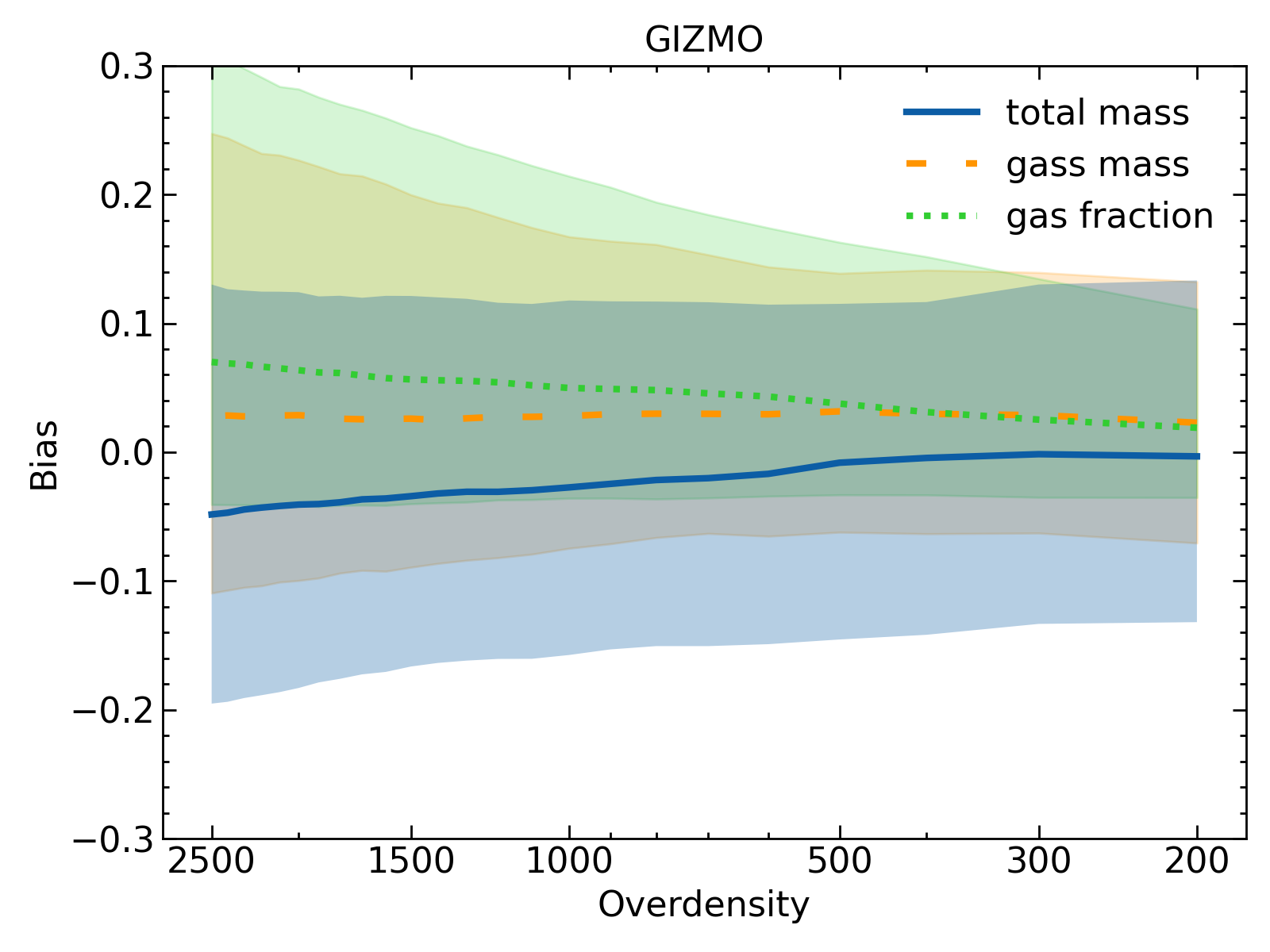}
    \caption{The median bias of \gadgetx{} (top panel) and \gizmo{} (bottom panel) clusters as a function of the overdensity for total mass (blue solid line), gas mass (red dashed line) and gas fraction (green dotted line). The shaded light blue, pink and green regions correspond to the $16^\text{th}-84^\text{th}$ percentiles for total, gas mass and gas fraction, respectively.}
    \label{fig:bias_mix_single}
\end{figure}

The most interesting results are those obtained by applying this network to test sets consisting of clusters from each run separately.
In the case of the test set of only \gadgetx{} clusters (top panel of fig.~\ref{fig:bias_mix_single}), the median bias of the predicted total mass profiles (blue solid line) is zero, and the scatter (blue shaded region) is also perfectly in accordance with what was achieved with the network trained with only \gadgetx{} clusters. On the other hand, regarding the gas mass profiles (orange dashed line), we observe a bias that slightly decreases to $\sim -5\%$ towards the cluster core whereas the scatter remains similar to that of the previous cases.
In the bottom panel of Fig.~\ref{fig:bias_mix_single} we show the results of the network applied to the \gizmo{} clusters. In this case, the median bias of the total mass profiles is zero up to $\Delta \leq 500$ and then it grows to about 5\% in the core. The scatter grows slightly but remains below 20\%. The gas mass profiles show a fairly constant bias around 3\%, moreover, for these profiles the scatter increases significantly in the inner part of the clusters, This might be caused by the strong AGN-feedback implemented in \gizmo{}.  In contrast, \gadgetx{} predictions are more stable in the centre. 

In conclusion, the approach of training the network with a mixture of clusters from different simulations gives accurate results in estimating the profiles of the two tests separately, showing the flexibility of our ML model.
 Differences, as might be expected, are observed for the gas mass profiles as well as in the innermost part of the clusters where the different baryon physics models have greater effects.


\bsp	
\label{lastpage}
\end{document}